\documentclass[letterpaper,twocolumn,10pt]{article}
\usepackage{usenix2021_SOUPS}
\usepackage{booktabs}
\usepackage[T1]{fontenc}
\usepackage[utf8]{inputenc}
\usepackage{textcomp} 
\usepackage{caption,subcaption}
\usepackage{graphicx}

\newcommand\enquote[1]{``\textit{#1}''}

\microtypecontext{spacing=nonfrench}

\begin{document}

\date{}

\title{\Large \bf \textit{A Fait Accompli?} An Empirical~Study into the Absence of Consent to Third-Party~Tracking in Android~Apps}

\def\plainauthor{Konrad Kollnig et al.}


\author{
{\rm Konrad Kollnig, Reuben Binns, Pierre Dewitte*, Max Van Kleek,}\\
{\rm Ge Wang, Daniel Omeiza, Helena Webb, Nigel Shadbolt}\\
Department of Computer Science, University of Oxford, UK\\
{*Centre for IT and IP Law, KU Leuven, Belgium}\\
firstname.lastname@(cs.ox.ac.uk | kuleuven.be)
} 

\maketitle
\thecopyright

\begin{abstract}
    Third-party tracking allows companies to collect users' behavioural data and track their activity across digital devices.
    This can put deep insights into users' private lives into the hands of strangers,
    and often happens without users' awareness or explicit consent.
    EU and UK data protection law, however, requires consent, both 1) to access and store information on users' devices and 2) to legitimate the processing of personal data as part of third-party tracking, as we analyse in this paper.

    This paper further investigates whether and to what extent consent is implemented in mobile apps. First, we analyse a representative sample of apps from the Google Play Store. We find that most apps engage in third-party tracking, but few obtained consent before doing so, indicating potentially widespread violations of EU and UK privacy law. Second, we examine the most common third-party tracking libraries in detail. While most acknowledge that they rely on app developers to obtain consent on their behalf, they typically fail to put in place robust measures to ensure this: disclosure of consent requirements is limited; default consent implementations are lacking; and compliance guidance is difficult to find, hard to read, and poorly maintained.
\end{abstract}

\begin{figure*}
    \centering
    \begin{subfigure}[t]{0.3\linewidth}
        \centering
        \includegraphics[width=\textwidth]{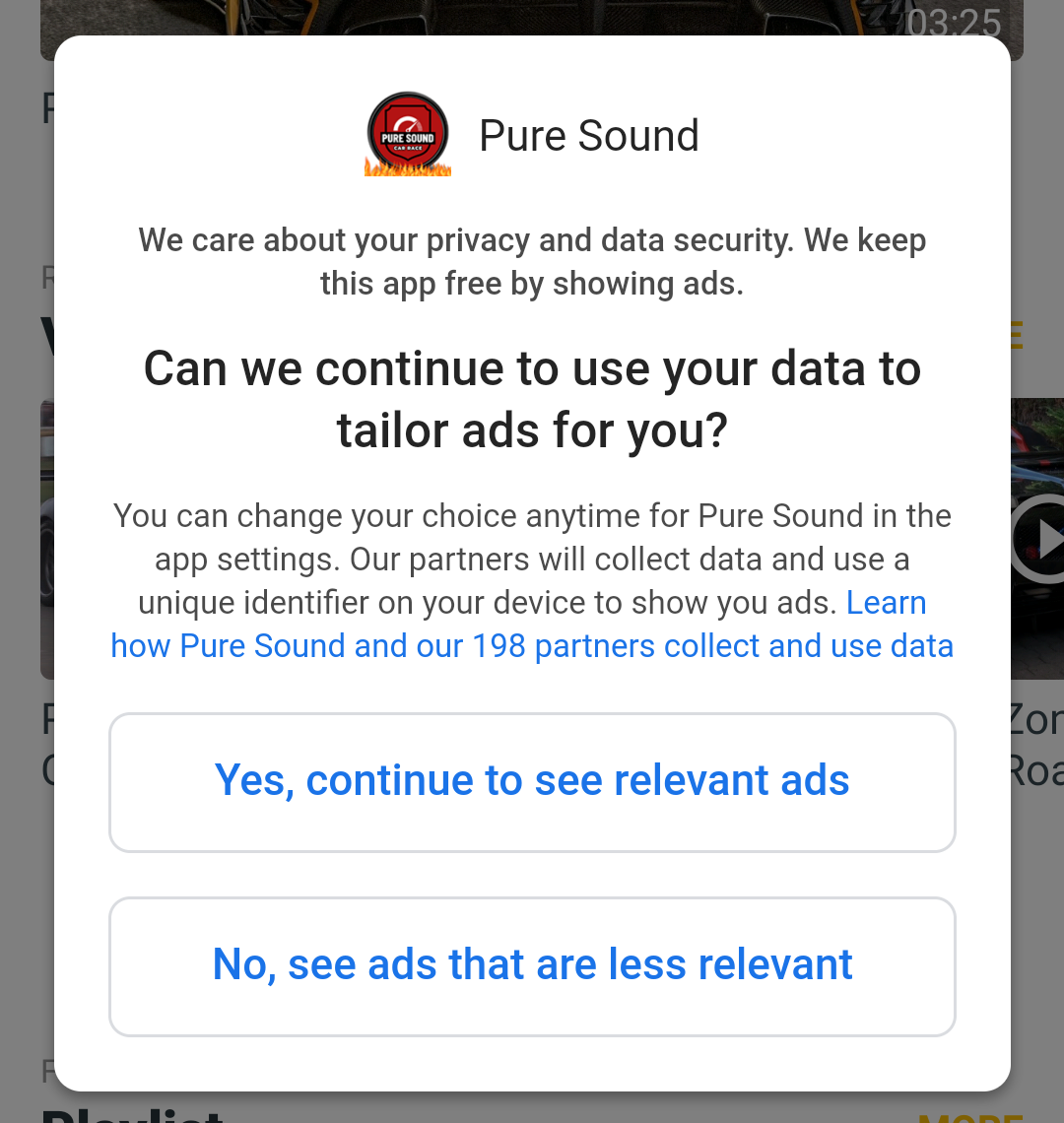}
        \caption{This app uses the Consent API developed by Google. The popup suggests that personal data may be shared with 199 companies before user consent is given (`continue').
        }
        \label{fig:google_consent}
    \end{subfigure}
    \hfill
    \begin{subfigure}[t]{0.3\linewidth}
        \centering
        \includegraphics[width=\textwidth]{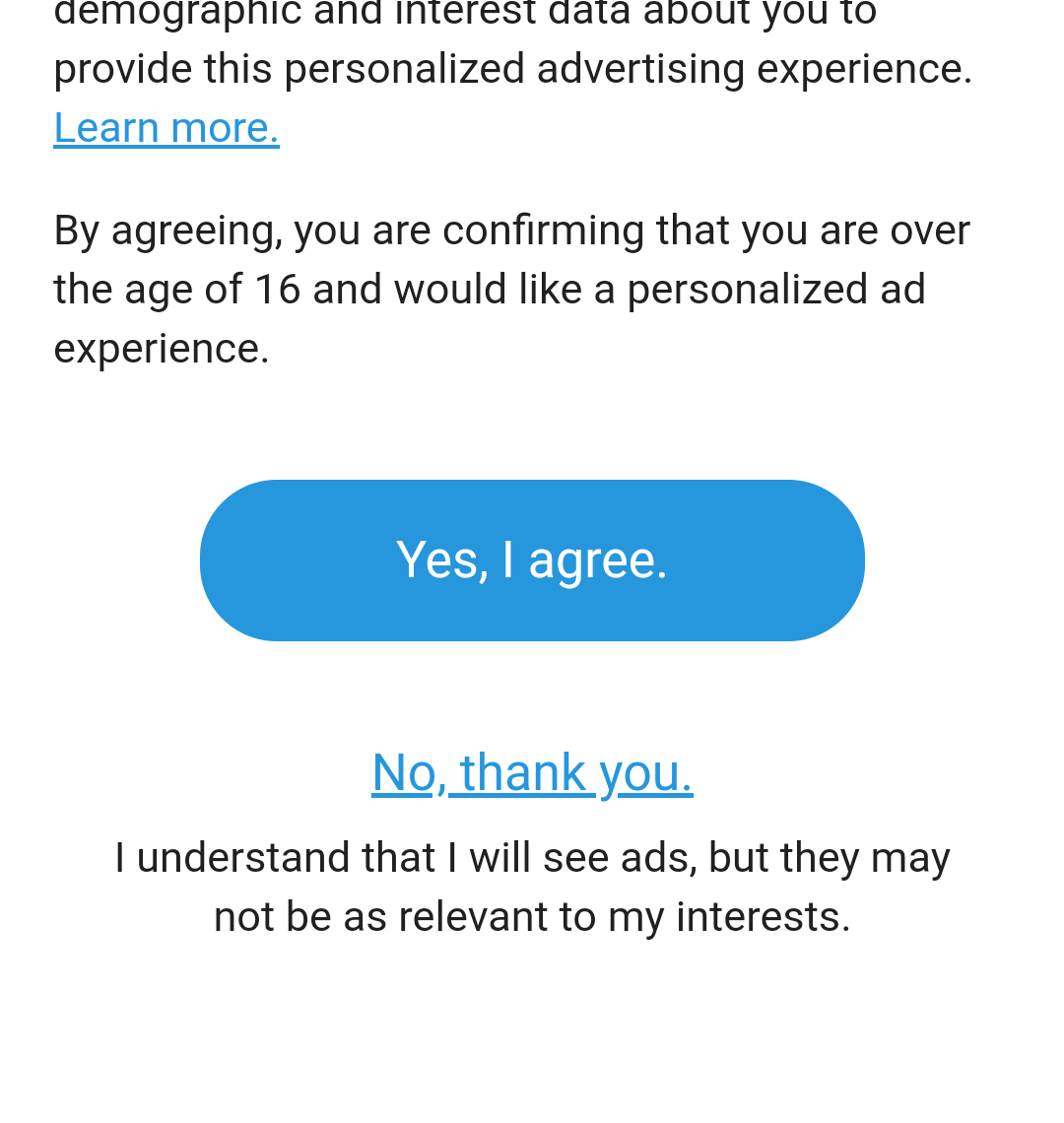}
        \caption{This app uses the consent implementation by Twitter MoPub. By declining, a user rejects a `personalized ad experience', but potentially not all app tracking.}
        \label{fig:mopub_consent}
    \end{subfigure}
    \hfill
    \begin{subfigure}[t]{0.3\linewidth}
        \centering
        \includegraphics[width=\textwidth]{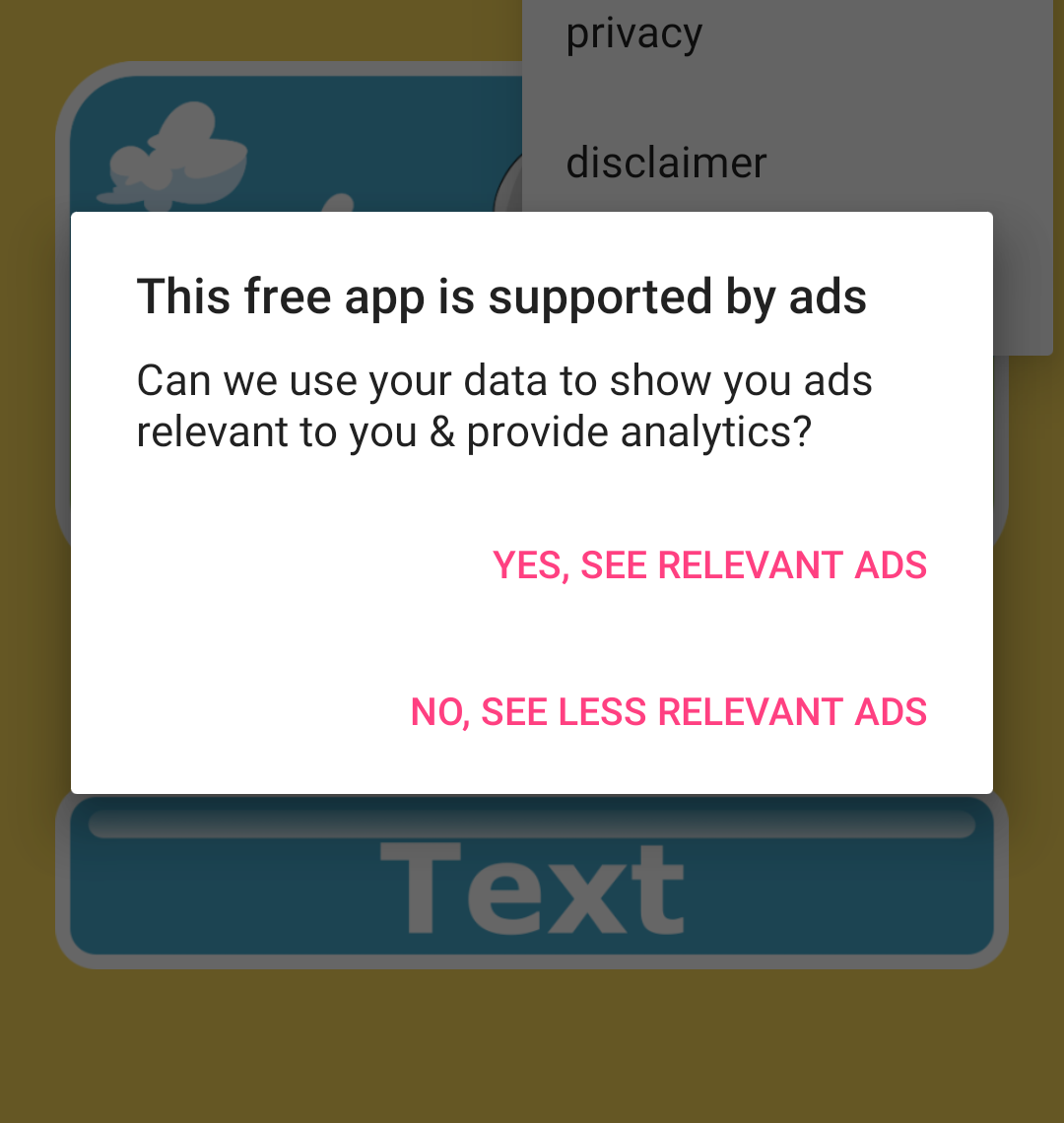}
        \caption{This app uses a custom consent solution. Consent is not granular. The answer options do not match the question. It is unclear if `No' rejects analytics.}
        \label{fig:custom_consent}
    \end{subfigure}
    \caption{While most apps on the Google Play Store use third-party tracking, only few apps allow users to refuse consent (less than $3.5\%$). The figure shows three common examples of these $3.5\%$ of apps.
    Since very few apps give users a genuine choice over tracking, our paper suggests widespread violations of EU and UK privacy law.
    }
    \label{fig:consent_ads}
\end{figure*}

\section{Introduction}
Third-party tracking, the deliberate collection, processing and sharing of users' behavioural data with third-party companies, has become widespread across both mobile app ecosystems~\cite{maps_2019,china_2018,binns_measuring_2018} and the web~\cite{binns_measuring_2018,papadogiannakis_user_2021}.
The use of third-party trackers benefits app developers in several ways, notably by providing analytics to improve user retention, and by enabling the placement of personalised advertising within apps, which often translates into a vital source of revenue for them~\cite{anirudhchi2021,mhaidli_we_2019}.
However, it also makes app developers dependent on privacy-invasive data practices that involve the processing of large amounts of personal data~\cite{greene_platform_2018, mhaidli_we_2019, fathaigh_mobile_2018}, with little awareness from users and app developers~\cite{datenschutzkonferenz_orientierungshilfe_2019, reyes_wont_2018, maps_2019, solove_privacy_2012}.
Data protection and privacy legislation such as the General Data Protection Regulation (GDPR)~\cite{gdpr} in the EU and the UK, and the Children's Online Privacy Protection Act (COPPA)~\cite{coppa} in the US, establish clear rules when it comes to the processing of personal data and provide additional safeguards when it comes to information relating to children.
As explained in Section~\ref{sec:consent_law}, consent is a necessary precondition for third-party tracking.

The implementation of consent in mobile apps has---since the end of 2020---sparked a fierce public battle between Apple and Facebook over tracking controls in iOS 14.5~\cite{ios1,ios2}.
To give users more control over their data, Apple has introduced an opt-in mechanism for the use of the Advertising Identifier (AdID)---similar to how apps currently request location or contacts access.
Facebook, like many other mobile advertising companies, is concerned that most users will not agree to tracking if asked more clearly and explicitly~\cite{ios3}; iOS users could already opt-out from the use of AdID, but were not explicitly asked by every app.
By comparison, Google does not currently offer users the option to prevent apps from accessing the AdID on Android in general, but intends to change this from `late 2021'~\cite{google_adid_change}.
The importance of consent aside, there exists little empirical evidence as to whether mobile apps implement any type of consent mechanisms before engaging in tracking.

Despite their crucial role within the software development life cycle, putting the blame of implementing consent incorrectly on app developers might be misguided. 
Many lack legal expertise, depend on the use of tracking software, and face limited negotiation power in the design of tracker software, which is usually developed by large, multinational companies~\cite{anirudhchi2021,balebako_privacy_2014,chitkara_does_2017,hadar_privacy_2018,mhaidli_we_2019}.
At the same time,
failure to implement appropriate consent mechanisms in software impacts individuals' choice over data collection and their informational self-determination, and may expose vulnerable groups---such as children---to disproportionate data collection. This underlines the need for robust privacy guarantees in code.

Driven by these observations, the present contribution aims to answer the following research questions:
\begin{enumerate}
    \item Do app developers need to obtain valid user consent before engaging in third-party tracking in the EU and UK? \textit{(consent requirements for tracking)}
    \item To what extent do apps engage in third-party tracking, and obtain valid user consent before doing so? \textit{(practices of app developers)}
    \item To what extent do third-party tracking companies encourage and support app developers to obtain consent as and where required? \textit{(practices of tracker companies)}
\end{enumerate}

\textbf{Contributions.} In answering these questions, this paper makes three contributions.
First, we clarify the role of consent in the regulatory framework applicable in the EU and the UK when it comes to the processing of personal data for third-party tracking.
Second, we provide empirical evidence as to a widespread absence of consent mechanisms to legitimise third-party tracking in 1,297 apps.
Third, we analyse the guidance provided by 13 commonly used tracker companies and assess whether they inform app developers about how to translate consent in code (see Figure~\ref{fig:notfound} and Table~\ref{tab:tracker_sdks}).

\textbf{Structure.} The rest of this paper is structured as follows. 
Section~\ref{sec:tech_background} reviews the relevant literature surrounding the concept of consent, app privacy analysis, and existing system-wide tracking controls for Android. 
Section~\ref{sec:consent_law} discusses the role of consent for third-party tracking in the EU and UK by drawing on the guidance issued by national Data Protection Authorities (DPAs).
Section~\ref{sec:consent_study} analyses the presence of consent for third-party tracking in 1,297 Android apps randomly sampled from the Google Play Store.
Section~\ref{sec:developers} reviews the guidance offered by tracker companies to app developers.
After discussing the limitations of our approach in Section~\ref{sec:empirical_limitations}, we turn to the discussion of our results in Section~\ref{sec:discussion} and our conclusions in Section~\ref{sec:conclusions}.

\section{Background}
\label{sec:tech_background}
In this section, we discuss previous relevant literature, covering the concept of consent, the empirical analysis of privacy in apps, and existing system-wide tracking controls for Android.
In particular, we highlight the limits of consent, and the dependence of end-users on the privacy options implemented by their apps and smartphone operating system.

\subsection{Promises and Limits of Consent}

Consent is a pillar of privacy and data protection law, in the US, EU, and many other jurisdictions and international frameworks. As an approach to privacy protection, consent is associated with the regime of \emph{notice \& choice}~\cite{solove_privacy_2012}.
For many data-processing activities, companies that want to process data from an individual must
\begin{enumerate}
    \item Adequately inform the individual  (\textit{Notice}), and
    \item Obtain consent from the individual (\textit{Choice}).
\end{enumerate}
These two fundamental requirements are often implemented in software through the provision of a privacy policy, accompanied by consent options for the end-user.


The limitations of the notice \& choice paradigm have been explored in a range of scholarship. 
Regarding \enquote{notice}, it has been documented that most people do not read privacy policies, and that when they try to, have difficulties understanding them~\cite{reidenberg_ambiguity_2016} and do not have enough time to read every such policy~\cite{mcdonald_cost_2008}.

Regarding \enquote{choice}, evidence suggests that many individuals struggle with privacy decisions in practice~\cite{shklovski_leakiness_2014, acquisti_nudging_2009}. 
The mismatch between stated and observed privacy preferences is known as the \enquote{privacy paradox}~\cite{norberg_privacy_2007}, although this so-called \enquote{paradox} may be best explained by structural forces that prevent alignment between values and behaviour\cite{solove2020myth,waldman2020cognitive}. 
Individuals often have no real choice but to accept certain data processing because some digital services---such as Facebook or Google---have become indispensable~\cite{bundeskartellamt_b6-2216_2019}. 
Even when offered genuine choice, individuals face ubiquitous tracking~\cite{binns_measuring_2018}, are tricked into consent~\cite{nouwens_dark_2020}, and do not get an adequate compensation in exchange for their data~\cite{competition_and_markets_authority_online_2020}.
Because of the limits to individual privacy management, various scholars argue that the regime of notice \& choice does not provide \emph{meaningful} ways for individuals to manage their privacy~\cite{solove_privacy_2012,barocas2009notice,bietti_consent_2020}.

Despite such limitations, consent remains a key component of many privacy and data protection regimes. 
For the purpose of this present contribution, we do not assume that consent is the only or best way to address privacy and data protection issues. 
Rather, we aim to investigate whether, in addition to all these problems and limitations, the basic process of consent itself is even being followed where it is currently required in the context of third-party tracking in apps.

\subsection{Analysing Privacy in Apps}

There is a vast range of previous literature that has analysed the privacy practices of mobile apps, and third-party tracking in particular.
Two main methods have emerged in the academic literature:
dynamic and static analysis.

\textbf{Dynamic analysis} executes an app, and analyses its run-time behaviour.
While early research analysed apps by modifying the operating system~\cite{enck_taintdroid_2010, agarwal_protectmyprivacy_2013}, recent work has focused on analysing apps' network traffic~\cite{ren_recon_2016,van_kleek_better_2017,reyes_wont_2018,privacyguard_vpn_2015,lumen_2018,antmonitor_2017,nomoads_2018,free_v_paid_2019,okoyomon_ridiculousness_2019}.

As for system modification, Enck et al. modified Android so that sensitive data flows through and off the smartphone could be monitored easily~\cite{enck_taintdroid_2010}. Agarwal and Hall modified iOS so that users were asked for consent to the usage of sensitive information by apps~\cite{agarwal_protectmyprivacy_2013}, before the introduction of run-time permissions by Apple in iOS~6.

As for network analysis, Ren et al. instrumented the VPN functionality of Android, iOS, and Windows Phone to expose leaks of personal data over the Internet~\cite{ren_recon_2016}.
Conducting a manual traffic analysis of 100 Google Play and 100 iOS apps, they found regular sharing of personal data in plain text, including device identifiers (47 iOS, 52 Google Play apps), user location (26 iOS, 14 Google Play apps), and user credentials (8 iOS, 7 Google Play apps).
Van Kleek et al. used dynamic analysis to expose unexpected data flows to users and design better privacy indicators for smartphones~\cite{van_kleek_better_2017}.
Reyes et al. used dynamic analysis to assess the compliance of children's apps with COPPA~\cite{reyes_wont_2018}, a US privacy law to protect children.
Having found that 73\% of studied children's apps transmit personal data over the Internet, they argued that none of these apps had obtained the required \enquote{verifiable parental consent} because their automated testing tool could trigger these network calls, and a child could likely do so as well.
Okoyomon et al. found widespread data transmissions in apps that were not disclosed in apps' privacy policies, and raised doubts about the efficacy of the notice \& choice regime~\cite{okoyomon_ridiculousness_2019} (as discussed in the previous section).

Dynamic analysis offers different advantages. It is relatively simple to do, largely device-independent, and can be used to monitor what data sharing actually takes place. It has, however, several limitations. The information gathered might be incomplete if not all code paths within the app involving potential data disclosures are run when the app is being analysed.
Further, network-based dynamic analysis may wrongly attribute system-level communications to a studied app, e.g. an Android device synchronising the Google Calendar in the background, or conducting a network connectivity check with Google servers. Network-based dynamic analysis remains nonetheless a versatile, reliable and practical approach.

\textbf{Static analysis} infers the behaviour of an app without the need for execution.
This process often relies on decompiling an app and analysing the retrieved program code~\cite{egele_pios_2011, han_comparing_2013}.
The main advantage of static analysis is that it enables the analysis of apps at a much larger scale (e.g. millions rather than hundreds) ~\cite{playdrone_2014, binns_measuring_2018, chen_following_2016,china_2018}.
As opposed to dynamic analysis, static analysis may require substantial computing resources and does not permit the direct observation of network traffic because apps are never run.

Egele et al. developed an iOS decompiler and analysed 1,407 iOS apps. They found that
55\% of those apps included third-party tracking libraries~\cite{egele_pios_2011}.
Viennot et al. analysed more than 1 million apps from the Google Play Store, and monitored the changing characteristics of apps over time~\cite{playdrone_2014}. They found a widespread presence of third-party tracking libraries in apps (including Google Ads in 35.73\% of apps, the Facebook SDK in 12.29\%, and Google Analytics in 10.28\%).
Similarly, Binns et al. found in analysing nearly 1 million Google Play apps that about 90\% may share data with Google, and 40\% with Facebook~\cite{binns_measuring_2018}.

The presence of consent to tracking in apps has received relatively little research attention; to the best of our knowledge, no large-scale studies in this area exist.
With static analysis, it is difficult to detect at what stage a user might give consent, because of the varied implementations of consent in app code.
However, network-based dynamic analysis makes this kind of consent analysis possible, and at a reasonable scale. We demonstrate this in Section~\ref{sec:consent_study}.

\subsection{Alternatives to In-App Consent}
\label{sec:tracking_tools}

Before turning to the legal analysis concerning when consent for third-party tracking within individual apps is required, it is worth considering the options users currently have to limit app tracking on Android at a system level. 
This is pertinent to our subsequent analysis because, if system-level controls were sufficient, the question of efficacy and compliance with individual app-level consent requirements might be redundant. 
The options for users fall into three categories: system settings, system modification, and system APIs.

\textbf{System Settings.}
The Android operating system offers users certain possibilities to limit unwanted data collection. 
Users can manage the types of data each app can access through \emph{permissions}.
This does not stop tracking, but blocks access to certain types of data, such as location.
A problem inherent to the permission approach is that trackers share permission access with the apps they come bundled with.
This means that, if a user allows location access to a maps app with integrated trackers, all these trackers have access as well.
This, in turn, might give users a false sense of security and control.
Google offers users the possibility to opt-out from personalised advertising.
If users choose to do so, apps are encouraged to cease using the system-wide \emph{Google Advertising Identifier} (AdID) for personalised advertising (although apps can continue to access the AdID).
Unlike iOS, Android does yet not offer the option to opt-out from analytics tracking using the AdID, or to prevent apps from accessing this unique user identifier. However, Google intends to change this from `late 2021'~\cite{google_adid_change}.

\textbf{System Modification.}
Since the early days of Android, many developers have set out to modify its functionality and implement better privacy protections. \emph{Custom ROMs} are modified versions of Android that replace the default operating system that comes pre-installed on Android smartphones. 
Popular examples are Lineage OS and GrapheneOS, which both try to reduce the dependency on Google on Android and increase user privacy. 
Another is TaintDroid, which monitors the flow of sensitive information through the system~\cite{enck_taintdroid_2010}.
A popular alternative to custom ROMs is \emph{rooting} devices by using exploits in the Android system to gain elevated access to the operating system, or by changing the bootloader of the Android system.
Rooting is a necessary prerequisite for many privacy-focused apps, including AdAway~\cite{adaway},
XPrivacy~\cite{xprivacy}, and
{AppWarden}~\cite{appwarden}.
System modification grants maximum control and flexibility regarding tracking, but requires a high level of technical expertise. 
It also relies on security vulnerabilities, often creating risks for (non-expert) users.

As a result, Google has recently begun to restrict attempts to modify Android by preventing custom ROMs from running apps using \emph{Google's Safety Net}.
This is meant to protect sensitive apps (e.g. banking apps) from running on unsafe devices, but is also used by other popular apps such as Pokemon GO and Snapchat~\cite{safetynet}. 
Some Internet outlets have declared the \enquote{end for Android rooting, [and] custom ROMs}~\cite{safetynet_change}.

\textbf{System APIs.}
Another alternative to system modification is to develop apps that build on the capabilities of Android's system APIs to detect and block network traffic related to tracking.
Such is possible without the need for system modification at the cost of more advanced functionality.
Popular apps in this category include
AdGuard (using a local VPN on the Android device)~\cite{adguard} and
DNS66 (changing the DNS settings of the Android device)~\cite{dns66}.
Another is NetGuard~\cite{netguard}, a  firewall that allows users to monitor network connections through a VPN, and to block certain domains manually.
All these tools block connections regardless of the actual content of the communications.
These content-agnostic approaches can lead to overblocking and break apps.

Alternative tools aim for more fine-grained protection by removing sensitive information from network requests, such as device identifiers or location data~\cite{privacyguard_vpn_2015,antmonitor_2017}.
Unfortunately, these content-based approaches rely on breaking secured network connections, and on installing a self-signed root certificate on the user's device. 
This practice was banned by Google with the introduction of Android 7 in 2016 because of the security risks it entails~\cite{google_ca}.
While these apps grant users the possibility to block tracking through system APIs, Google does not allow them on the Play Store~\cite{playstore_interference}. 
Instead, users must sideload them onto their device from alternative sources, such as GitHub and F-Droid.

In conclusion, while there exists a wide array of options for end-users to reduce tracking, none of them can provide the granularity of consent implemented inside each individual app.
Many of the existing tools require a high level of technical expertise, including root access or modifications to the operating system, and are therefore unsuitable for non-expert users.
This makes many users dependent on the privacy solutions offered by apps themselves and their operating systems.

\section{When is Consent to Tracking required?}
\label{sec:consent_law}
In this section, we analyse whether consent is a prerequisite for third-party tracking under EU and UK law, as well as its role under the Google Play Store policy. 
We focus on these jurisdictions as they have relatively stringent and specific rules on consent and third-party tracking. 
While similar rules exist in other jurisdictions (such as the COPPA in the US, which requires parental consent for tracking), recent regulatory actions and rich guidance issued by European regulators offer an ideal setting for a large-scale analysis.

Two main legal instruments are relevant to the issue of consent to third-party tracking on mobile apps: the GDPR and the ePrivacy Directive\footnote{It is worth noting that the ePrivacy Directive is currently under revision. A change to the current regulatory requirements in practice is expected no earlier than in two years due to the nature of the EU legislative process.}.

\subsection{GDPR and the Need for a Lawful Ground}

Applicable since 25 May 2018, the GDPR grants users various rights over their \emph{personal data}.
It also imposes a wide array of obligations on so-called \emph{controllers}, paired with high fines for non-compliance.
One of the cornerstones of the legislative reform is the concept of \emph{Data Protection by Design}; this obliges controllers to implement appropriate technical and organisational measures to ensure and demonstrate compliance with all rules stemming from the Regulation throughout the entire personal data processing life cycle (Article 24(1) and 25(1) GDPR).
All companies operating in the EU or UK must follow the rules set out in the GDPR.\footnote{More specifically, all companies based in the EU and UK, as well as companies monitoring the behaviour of, or offering goods and services to, individuals located in the EU and UK, fall within the territorial scope of application of the GDPR (Article 3 GDPR).}
Compliance with the GDPR is monitored and enforced by the \emph{Data Protection Authorities} (DPAs) instituted in each EU Member State and in the UK.
If a controller fails to comply, DPAs have the power to impose fines that can go up to €20 million or 4\% of the company's worldwide annual turnover, whichever is higher.

For the purpose of this paper, we assume that app developers qualify as \emph{controllers}. 
In other words, that they \enquote{determine the purposes and the means of the processing of personal data} (Article 4(7) GDPR). 
While this might well be the case when the company actually processing the personal data at stake is also in charge of the development of the app, it is important to highlight that controllership does not always end up on their shoulders. This is the case, for instance, when a company outsources the development of its app to an external team of software developers working on the basis of clear-cut specifications and requirements, in which case the latter is likely to be considered as a \emph{processor} (Article 4(8) GDPR) or a \emph{third party} (Article 4(10) GDPR). 

If app developers want to collect personal data for whatever purpose, they need to rely on one of the six \emph{lawful grounds} listed in Article 6(1) GDPR.
Only two usually apply in the context of mobile apps, namely: \textit{consent} and \textit{legitimate interests}\footnote{The remaining four lawful grounds listed in Article 6(1) GDPR being the fulfilment of a \textit{contract}, a \textit{legal obligation}, the data subject's \textit{vital interests}, and the performance of a \textit{public task}.}.
On the one hand, and as specified in Article 4(11) GDPR, a valid \textit{consent} is
\begin{quote}
    \textit{any freely given, specific, informed and unambiguous indication of the data subject's wishes by which he or she, by a statement or by a clear affirmative action, signifies agreement to the processing of personal data relating to him or her}
\end{quote}
As recently clarified by the Court of Justice of the European Union, this bans the use of pre-ticked boxes to gather consent~\cite{ecj_consent}.
\textit{Legitimate interests}, on the other hand, is a viable alternative to consent but requires a careful balancing exercise between the controller's interests in processing the personal data and the data subjects' interests and fundamental rights (Article 6(1)f GDPR)~\cite{wp29_li}.
The other guarantees stemming from the GDPR (including transparency, security, purpose and storage limitation, and data minimisation) remain applicable regardless of the lawful ground used to legitimise the processing.

\textbf{Consent for High-Risk Data Processing.}
While the controller's legitimate interests could potentially be a viable option for legitimising third-party tracking on mobile apps, this processing is likely to qualify as a \textit{high-risk data processing activity}.\footnote{The Article 29 Working Party---an EU body
to provide guidance on data protection law (now the European Data Protection Board)---has listed the 9 features commonly found in such high-risk activities, namely: 1) Evaluation or scoring, 2) Automated-decision making with legal or similar significant effect, 3) Systematic monitoring, 4) Sensitive data or data of a highly personal nature, 5) Data processed on a large scale, 6) Matching or combining datasets, 7) Data concerning vulnerable data subjects, 8) Innovative use or applying new technological or organisational solutions, and 9) Prevention of data subjects from exercising a right or using a service or a contract.~\cite{highrisk}} 
Features of third-party tracking that indicate such high-risk processing include the use of \enquote{evaluation or scoring}, \enquote{systematic monitoring}, \enquote{data processed on a large scale}, \enquote{data concerning vulnerable data subjects}, or \enquote{innovative use or applying new technological or organisational solutions}. 
Some of these features undoubtedly apply to third-party tracking, since tracking companies usually engage in large-scale data collection, at a high-frequency, across different services and devices, with limited user awareness.

The Information Commissioner's Office (ICO)---the UK's DPA---discourages the use of legitimate interest for high-risk activities and recommends that controllers
instead rely on another lawful ground such as consent~\cite{ico_hrisk}.
Similarly, after having analysed the case of tracking deployed on a webshop selling medical and cosmetic products~\cite{datenschutzkonferenz_orientierungshilfe_2019}, the German DPA came to the conclusion that the average website visitor could not reasonably expect tracking of their online activities to take place, especially when it operates across devices and services.
In that case, it argued, the website visitor is not in a position to avoid the data collection.
These are the concrete manifestations of the balancing exercise required by Article 6(1)f.
All in all, the above-mentioned considerations disqualify the use of the controllers' legitimate interests as an appropriate lawful ground to legitimise third-party tracking in mobile apps.

\subsection{ePrivacy and the Need for Consent for Local Storage of and Access to Data}
\label{sec:eprivacy}
In addition to the GDPR, the ePrivacy Directive also applies to third-party tracking. 
This is a \emph{lex specialis}, meaning that, when both the ePrivacy Directive and the GDPR apply in a given situation, the rules of the former will override the latter. 
This is the case for third-party tracking, since Article 5(3) of the ePrivacy Directive specifically requires consent for accessing or storing non-technically necessary data on a user's device.
It is widely accepted, and reflected in DPAs' guidance, that most tracking activities are not technically necessary, and therefore require consent to store data on a user's device~\cite{ico_cookies}.
So, if tracker software involves accessing or saving information on a user's smartphone---as third-party trackers typically do on a regular basis---this requires prior consent.
As a result, while consent was already the most reasonable option under the GDPR, it becomes the only viable one when combining both regulatory frameworks.

As stated above, the GDPR provides a range of possible lawful grounds of which consent is just one; however, Article 5(3) of the ePrivacy Directive specifically requires consent for accessing or storing non-technically necessary data on a user's device.
As a consequence, any further processing by the third party which is not covered by the initial consent interaction would usually require the third party to obtain fresh consent from the data subject.

Recent guidance and enforcement action from various DPAs have also demonstrated how the GDPR and the ePrivacy requirements apply to situations where consent is the basis for processing by one controller, and when that data is provided to another controller for further processing. 
Article 7(1) of the GDPR requires that, where consent is the lawful ground, the controller must be able to \emph{demonstrate} that the data subject has consented.
The ICO's guidance states that third-party services should not only include contractual obligations with first parties to ensure valid consent is obtained, but \enquote{may need to take further steps, such as ensuring that the consents were validly obtained}~\cite{ico_cookies2}. 
It notes that, while the process of getting consent for third-party services \enquote{is more complex}, \enquote{everyone has a part to play}~\cite{ico_cookies2}.
The responsibility of third parties has been further illustrated in an enforcement action by the CNIL (the French DPA), against Vectaury, a third-party tracking company~\cite{cnil_vectuary}. 
This showed how the validity of consent obtained by an app developer is not \enquote{transitive}, i.e. does not carry over to the third party. 
If a first party obtains consent \enquote{on behalf} of a third party, according to a contract between the two, the third party is \emph{still} under the obligation to verify that the consent is valid.

To summarise the implications of GDPR and ePrivacy in the context of third-party tracking: consent is typically required for access to and storage of data on the end-user's device. Even if that consent is facilitated by the first party, third parties must also be able to demonstrate the validity of the consent for their processing to be lawful on that basis.

\subsection{Requirements of the Google Play Store}
In addition to EU and UK privacy law, Google imposes a layer of contractual obligations that apps must comply with.
These policies apply worldwide---so beyond the jurisdiction of the EU and UK---and might oblige all app developers to implement adequate mechanisms to gather consent for third-party tracking.
Google's \emph{Developer Content Policy} highlights that in-app disclosure and consent might need to be implemented when \enquote{data collection occurs in the background of your app}~\cite{google_consent2}.
The Developer Content Policy also requires that developers abide by all applicable laws.
It is unclear how strictly compliance with these policies---and in particular with all applicable laws---is verified and enforced by Google.

\section{Tracking in Apps Before and After Consent}
\label{sec:consent_study}
The previous section established that third-party tracking in apps requires valid user consent under the EU and UK regulatory framework. Despite these legal obligations, it yet not clear how and whether consent is realised in practice. In order to examine the extent to which regulation around consent is implemented in practice, we conducted two studies---Study 1 (in this section) to see how consent is implemented in a representative sample of Google Play apps, and Study 2 (in the following Section~\ref{sec:developers}) to examine how app developers were supported and encouraged to implement consent by the providers of tracker libraries. 

\subsection{Methodology}

We studied a representative sample of 1,297 free Android apps from the UK Google Play Store.
This sample was chosen randomly (through random sampling without replacement) from a large set of 1.63 million apps found on the Google Play Store
between December 2019 and May 2020 to understand the practices across the breadth of app developers.
We explored the presence of apps on the app store by interfacing with Google Play's search function, similar to previous research~\cite{playdrone_2014}.
The selected apps were run on a Google Pixel 4 with Android 10.
Each app was installed, run for 15 seconds, and then uninstalled.
We did not interact with the app during this time, to record what companies the app contacts before the user can be informed about data collection, let alone give consent.
During app execution, we recorded the network traffic of all tested apps with the popular NetGuard traffic analysis tool~\cite{netguard}.
We did not include any background network traffic by other apps, such as the Google Play Services.
For apps that showed full-screen popup ads, we closed such popups, and took note of the presence of display advertising.
We assessed whether each contacted domain could be used for tracking
and, if so, to what tracking company it belonged,
using a combination of the App X-Ray~\cite{binns_third_2018} and \texttt{Disconnect.me}~\cite{disconnect} tracker databases.
15 seconds after having installed the app, we took a screenshot for further analysis, and uninstalled it.

We inspected the screenshots for any form of display advertising, privacy notice or consent.
We took note of any display advertising (such as banner and popups advertising) observed.
We classified any form of information about data practices as a privacy notice, and any \emph{affirmative} user agreement to data practices as consent.
While this definition of consent is arguably less strict than what is required under EU and UK law, this was a deliberate choice to increase the objectivity of our classification, and provide an upper bound on compliance with EU and UK consent requirements.
We then re-installed and ran those apps that asked for consent, granted consent, and repeated the network capture and analysis steps above, i.e.~monitoring network connections for 15 seconds, followed by a screenshot, and finally, removed the app once again.

\begin{table}
    \centering
    \begin{tabular}{@{}lll@{}}
        \toprule
        Hosts & Company & Apps\\
        \midrule
        \url{adservice.google.com} & Alphabet & 19.7\%\\
        \url{tpc.googlesyndication.com} & Alphabet & 17.2\%\\
        \url{lh3.googleusercontent.com} & Alphabet & 14.2\%\\
        \url{android.googleapis.com} & Alphabet & 12.9\%\\
        \url{csi.gstatic.com} & Alphabet & 11.6\%\\
        \url{googleads.g.doubleclick.net} & Alphabet & 10.3\%\\
        \url{ade.googlesyndication.com} & Alphabet & 9.7\%\\
        \url{connectivitycheck.gstatic.com} & Alphabet & 9.5\%\\
        \url{config.uca.cloud.unity3d.com} & Unity & 7.5\%\\
        \url{ajax.googleapis.com} & Alphabet & 6.9\%\\
        \url{api.uca.cloud.unity3d.com} & Unity & 6.8\%\\
        \url{android.clients.google.com} & Alphabet & 6.7\%\\
        \url{gstatic.com} & Alphabet & 5.8\%\\
        \url{graph.facebook.com} & Facebook & 5.5\%\\
        \bottomrule
    \end{tabular}
    \caption{Top contacted tracker domains by 1,201 randomly sampled apps from the Google Play Store, at launch, before any interaction with the apps.}
    \label{tab:domains}
\end{table}

\hypertarget{results-1}{%
\subsection{Results}\label{results-1}}

Of the 1,297 apps, 96 did not show a working user interface. 
Some apps did not start or showed to be discontinued. 
Other apps did not provide a user interface at all, such as widgets and Android themes. 
We therefore only considered the remaining 1,201 apps.
909 apps (76\%) were last updated after the GDPR became applicable on 25 May 2018.\footnote{It is worth noting, however, that both the need for a lawful ground---an obligation under Directive 95/46---and the consent requirement for access to and storing on terminal equipment---an obligation under the ePrivacy Directive---were already applicable before 25 May 2018. The latter has merely provided clarification on the conditions for consent to be valid.} 
On average, the considered apps were released in August 2018 and last updated in December 2018. All apps were tested in August 2020, within a single 24-hour time frame.

\textbf{Widespread tracker use.} 
Apps contacted an average of 4.7 hosts each at launch, prior to any user interaction. 
A majority of such apps (856, 71.3\%) contacted known tracker hosts. 
On average, apps contacted 2.9 tracker hosts each, with a standard deviation of 3.5. The top 10\% of apps contacted at least 7 distinct hosts each, while the bottom 10\% contacted none.
Alphabet, the parent company of Google, was the most commonly contacted company (from 58.6\% of apps), followed by Facebook~(8.2\%), Unity~(8.2\%), One Signal~(5.6\%), and Verizon~(2.9\%).
Apps that we observed showing display ads contacted a significantly higher number of tracker hosts (on average 6.0 with ads vs 2.2 without).

\textbf{Dominance of Google services.} 
The 9 most commonly contacted domains all belong to Google; the top 2 domains are part of Google's advertising business (\url{adservice.google.com}, linked to Google's Consent API, and \url{tpc.googlesyndication.com}, belonging to Google's real-time advertisement bidding service). 
704 apps (58.6\%) contacted at least one Google domain; the top (Google) domain was contacted by 236 apps (19.7\%). Such breadth and variation is reflective of the corresponding variety of services that Google offers for Android developers, including an ad network (Google AdMob), an ad exchange (Google Ad Manager, formerly known as DoubleClick), and various other services. Domains by other tracker companies, such as Unity and Facebook, were contacted less frequently by apps (see Table~\ref{tab:domains}).

Google's tracking was also observed to be deeply integrated into the Android operating system.
It has been known that the Google Play Services app---required to access basic Google services, including the Google Play Store---is involved in Google's analytics services~\cite{microg}.
In our network analysis, this app seemed to bundle analytics traffic of other apps and send this information to Google in the background with a time delay.
Without access to encrypted network traffic (as explained in Section~\ref{sec:tracking_tools}), this makes it impossible to attribute network traffic to individual apps from our sample, when such network traffic could also be related to other system apps (some of which, such as the Google Phone app, use Google Analytics tracking themselves).
As a consequence, we are likely under-reporting the number of apps that share data with Google, since we only report network traffic that could be clearly attributed.

\textbf{Consent to tracking is widely absent.}
Only 9.9\% of apps asked the user for consent. 
Apps that did so contacted a larger number of tracker hosts than those that did not (3.7 with consent vs 2.8 that did not).
A slightly larger fraction (12.2\% of all apps), informed the user to some extent about their privacy practices; apps in this category also contacted a larger number of trackers than those that did not (3.6 that informed vs 2.8 that did not).
19.1\% of apps that did not ask for consent showed ads, compared to only 2.5\% of apps that asked for consent. Once consent was granted, the apps contacted an average of 4.2 tracker hosts (higher than the 3.7 before granting consent, and the 2.8 for apps without any consent flows).

\textbf{Consent limited to using or not using an app.}
Most apps that ask for consent force users into granting it.
For instance, 43.7\% of apps asking for consent only provided a single choice, e.g.~a button entitled \enquote{Accept Policy and Use App} or obligatory check boxes with no alternative. 
A further 20.2\% of apps allowed users to give or refuse consent, but exited immediately on refusal, thus providing a \textit{Hobson's choice}.
Only 42 of the apps that implemented consent (comprising a mere 3.5\% of all apps) gave users a genuine choice to refuse consent. 
However, those apps had some of the highest numbers of tracker hosts, and contacted an average of 5.2 on launch.
Among these apps, if consent was granted, the number of tracker hosts contacted increased to 8.1, but, interestingly, an increase was also observed even if data tracking was opted-out (from the pre-consent 5.2 to 7.5 post-opt-out).

\textbf{Consent limited to the personalisation of ads.}
Consent was often limited to an opt-out from personalised ads. 
37 of the 42 apps that implement a genuine choice to refuse consent restrict this choice to limiting personalised advertising; such choice might make some users wrongly assume that refusing to see personalised ads prevents all tracking (see Figure~\ref{fig:consent_ads} for some common examples). 
We observed that 23 of these 37 apps (62\%; 1.9\% overall) used Google's Consent API~\cite{google_consent_api}, a toolkit provided by Google for retrieving consent to personalised ads (particularly when multiple ad networks are used). 
None of the apps using the Google Consent API, however, ended up asking users to agree to further tracking activities, such as analytics.
Only 4 apps provided the option to refuse analytics; all 4 of these did so in addition to providing the option to opt-out of personalised advertising. 
One further app in our sample requested consent to process health data.

\section{Support and Guidance from Trackers}
\label{sec:developers}
\begin{table*}
\centering
\begin{tabular}{@{}lllllll@{}}
\toprule
Tracker &
  Apps &
  \begin{tabular}[c]{@{}l@{}}Expects consent\\ (in EU / UK)\end{tabular} &
  \begin{tabular}[c]{@{}l@{}}Implements consent\\ (by default)\end{tabular} &
  \begin{tabular}[c]{@{}l@{}}Mentions consent\\ (in implementation guide)\end{tabular} & \begin{tabular}[c]{@{}l@{}}Discloses local\\ data storage\end{tabular} \\ \midrule
\textbf{Google Analytics}        & 50\% & Yes & No  & No  & Yes \\
Google AdMob                        & 45\% & Yes & No  & Yes & Yes  \\
\textbf{Google Crashlytics}      & 29\% & Yes & No  & No  & Yes  \\
\textbf{Facebook App Events}              & 20\% & Yes & No  & No  & ?  \\
\textbf{Google Tag Manager}      & 19\% & Yes & No  & No  & Yes  \\
Facebook Ads                  & 14\% & Yes & Yes* & No & ? \\
\textbf{Flurry}            & 9\%  & Yes & No  & No & ?   \\
Unity Ads                      & 8\%  & Yes & Yes & No & Yes \\
Inmobi                         & 8\%  & Yes & No  & Yes & ?  \\
Twitter MoPub                  & 6\%  & Yes & Yes & No & Yes \\
AppLovin         & 6\%  & No  & No  & No & ?   \\
AppsFlyer               & 5\%  & ?  & No  & Yes & ?  \\
\textbf{OneSignal}          & 4\%  & Yes & No  & No & Yes   \\ \bottomrule
\end{tabular}
\caption{Consent requirements and implementation for 13 commonly used Android trackers. App shares according to the Exodus Privacy Project~\cite{exodus-stats}. The \textbf{trackers in bold} require consent, but do neither implement such by default nor mention the need to do so in their implementation guides. ?: We did not find any information. *:~Facebook opts-in users by default to their personalised advertising, unless they disable this behaviour from their Facebook settings or do not use the Facebook app.}
\label{tab:tracker_sdks}
\end{table*}
The previous section found a widespread absence of consent to third-party tracking in apps. 
As explained in Section~\ref{sec:consent_law}, both first and third parties have a part to play in facilitating valid consent, and third parties need to take steps to ensure consent obtained by first parties is valid.
At the same time, it has been reported that many app developers
believe the responsibility of tackling risks related to ad tracking lie with the third-party companies~\cite{mhaidli_we_2019}, and need clear guidance regarding app privacy~\cite{balebako_privacy_2014}.
In this section, we assess the efforts, that providers of tracker libraries make, to encourage and support app developers in implementing a valid consent mechanism.
We focus on the most common libraries so as to understand the current practices across the tracking industry.

\subsection{Methodology}

Our qualitative analysis focuses on the 13 most common tracker companies on Android (according to~\cite{exodus-stats}), and three types of document that each of them provides: 1) a step-by-step implementation guide, 2) a privacy policy, and 3) further publicly available documentation. 
While there may be other ways in which providers of tracking libraries support app developers to facilitate valid consent, we reason that these are the standard means by which such support would be provided. 
Step-by-step implementation guides serve as a primary resource for app developers and summarise the essential steps of implementing a tracker library in code.
Since the implementation of consent must be done in code, consent implementation is one essential step for those trackers that require consent.

In assessing this documentation, we assume the perspective of an app developer who is motivated to comply with any explicit requirements mentioned by the tracker provider, and to follow their instructions as to how to do so, but lacks in-depth knowledge about how the GDPR and ePrivacy Directive apply to their use of a given third-party tracking software \cite{hadar_privacy_2018}. 
We also assume that app developers are likely to read documentation only so far as necessary to make the third-party library functional, often through trial-and-error~\cite{lawrance2010programmers,kelleher2019towards}, and stop studying other resources once the tracker implementation is functional, since they are often pressured by time and economic constraints~\cite{devs2016,anirudhchi2021,mhaidli_we_2019}.

\subsection{Results}

Our results are summarised in Table~\ref{tab:tracker_sdks}.
We detail our main findings in the following paragraphs.

\textbf{Most trackers are unclear about their use of local storage.} 
Whether a tracker accesses and/or stores information on a user's device is essential in determining the need to implement consent, as explained in Section~\ref{sec:eprivacy}.
As such, we would expect to find information stating whether or not access and/or storage takes place as part of the standard operation of the tracker.
However, we did not find such information for 6 out of 13 trackers. 
For the others, this information was difficult to find.
AppsFlyer rightly states in its online documentation that \enquote{there are no cookies for mobile apps or devices}~\cite{appsflyer_cookies}.
While this is true from a technical perspective, EU and UK law do not differentiate between cookies and other information saved on a user's device.
Crucially, we did not find any tracker stating \emph{not} to save information on a user's device. 
In the absence of such a denial, app developers would run the risk of assuming they do not need to obtain consent for data accessed and/or stored by the tracker.

\textbf{Most trackers expect app developers to obtain consent.}
Despite being unclear about their use of local storage, a closer inspection of the tracker policies and documentation found that most trackers instruct developers to request consent from EU users (11 out of 13). 
AppLovin is an exception, but does require consent if developers want to show personalised ads (which tend to be more lucrative than contextual ads).
For AppsFlyer, we could not find any information regarding the need to ask users for consent.
The need to ask for consent was sometimes difficult to find, and required a careful reading of the policies and documentation provided.
Some developers are bound to overlook this, and unnecessarily compromise on the users' right to choose over tracking.

\textbf{Few trackers implement consent by default.}
We further inspected whether tracker libraries provide their own consent implementation.
If they do, an app developer would not need to make any further modification to the app code.
However, only a minority of tracker libraries (3 out of 13) integrates an implementation of user consent by default, and none of the five most common trackers do so.
Unity Ads and Twitter MoPub provide consent flows that are automatically shown, without further action by the app developer.
Facebook Ads only shows ads, if the app user 1) has agreed to personalised ads in their Facebook account settings, and 2) uses the Facebook app on their phone.
However, Facebooks opts-in users by default to their personalised advertising, unless they disable this behaviour from their Facebook settings (checked 14 February 2021).
While Google AdMob provides a consent library, this is not implemented by default. Indeed, Google AdMob expects the app developer to retrieve consent from the user, but shows personalised ads even if the developer does not implement their consent library.

\textbf{Limited disclosure of consent requirements in step-by-step guides.}
We find that 3 out of 13 tracker libraries disclose the potential need for consent in their step-by-step implementation guides.
This is despite 11 out of 13 trackers mentioning the need to implement consent in other places of their online documentation.
Google AdMob mentions the need to retrieve consent amongst other \enquote{examples of actions that might be needed prior to initialization}~\cite{admob_quickstart} of AdMob.
Inmobi points out that developers need to \enquote{obtain appropriate consent from the user before making ad requests to InMobi for Europe}~\cite{inmobi} in the Section on 
\enquote{Initializing the SDK}.
AppsFlyer offers developers to \enquote{postpone start [of the tracker library] until you receive user consent due to GDPR or CCPA requirements, etc.}~\cite{appsflyer} in Section 3.4 on \enquote{Delay SDK initialization}.
It is not clear from these three implementation guides what other reasons are to \enquote{delay initialisation} beyond legal compliance, and why this is not clarified.
At least 6 out of 13 trackers require consent, but neither implement such by default nor inform app developers of the need to do so in the implementation guides.
If AppLovin needs consent (despite not stating to do so, but as suggested by our legal analysis in Section~\ref{sec:consent_law}), this figure would increase to 7 out of 13 trackers.

\begin{figure*}
    \centering
    \begin{subfigure}[t]{0.3\linewidth}
        \centering
        \includegraphics[width=\textwidth]{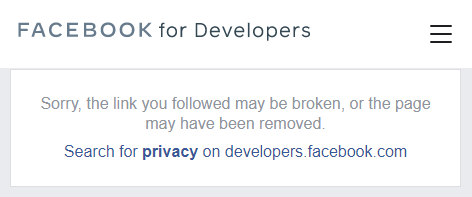}
        \caption{Facebook links to a page that is supposed to explain how to implement consent in practice.}
        \label{fig:facebook_guide}
    \end{subfigure}
    \hfill
    \begin{subfigure}[t]{0.3\linewidth}
        \centering
        \includegraphics[width=\textwidth]{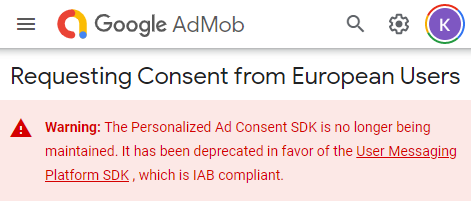}
        \caption{Google AdMob links to an outdated library, creating unnecessary friction for consent implementation.}
        \label{fig:admob_guide}
    \end{subfigure}
    \hfill
    \begin{subfigure}[t]{0.3\linewidth}
        \centering
        \includegraphics[width=\textwidth]{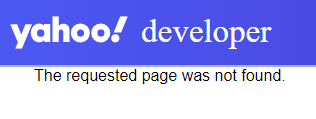}
        \caption{Flurry links to a broken GDPR guide.}
        \label{fig:flurry_guide}
    \end{subfigure}
    \caption{Many trackers provide information on what developers need to know to implement consent. These guides are often difficult to find, hard to read, and poorly maintained. 3 out of 13 common trackers linked to unmaintained or broken pages.}
    \label{fig:notfound}
\end{figure*}

\textbf{Compliance guidance: often provided, but sometimes difficult to find, hard to read, and poorly maintained.}
Many tracker companies provide additional information on GDPR compliance and consent implementation on a separate website as part of their online documentation. 
We found some compliance guidance (with varying levels of detail) for all trackers except the Google Tag Manager.
Excluding the 3 trackers implementing consent by default, a developer needs an average of 1.56 clicks to reach these compliance guides.
For AppLovin, a developer must click \enquote{Help Center}, then \enquote{Getting started \& FAQ}, and lastly \enquote{User opt-in/opt-out in the AppsFlyer SDK}.
Facebook required developers to click \enquote{Best Practices Guide} and then \enquote{GDPR Compliance guide}.
While this GDPR compliance guide provides some guidance on the implementation of consent, the link to Facebook's \enquote{consent guide} with practical examples of how to implement consent was broken.
Also, the framing as \enquote{Best Practices} suggests optionality of legal compliance.
For OneSignal, developers must first click \enquote{Data and Security Questions} and then \enquote{Handling Personal Data}.

The compliance guides (excluding code fragments) reached a mean Flesch readability score~\cite{flesch_new_1948} of 41.8, as compared to 50.6 for the step-by-step implementation guides (where 100 means \enquote{very easy}, and 0 \enquote{very difficult} to read).
Both the implementation and compliance guides are \enquote{difficult} to read, with the compliance guides somewhat more so.
For 3 of the 13 trackers, we were directed to broken or outdated links (see Figure~\ref{fig:notfound}). 
Google AdMob linked to an outdated consent strategy, while the Facebook SDK and Flurry linked to non-existing pages (returning \texttt{404} errors). 
We found other pages with compliance information for each of these trackers, but broken guidance can act as a deterrent for developers who want to implement consent and follow their legal obligations.
However, while this paper was under review, the broken links in the documentation of the Flurry and Facebook trackers were fixed.

\section{Limitations}
\label{sec:empirical_limitations}
It is important to acknowledge some limitations of our methodology.
Our analysis in Section~\ref{sec:consent_study} used dynamic analysis, and not all tracking might be detected.
We only inspected network traffic before and shortly after consent was given.
Apps might therefore conduct more tracking during prolonged app use.
Besides, we only reported the network traffic that could be clearly attributed to one of the apps we studied, potentially leading to under-reporting of the extent of Google's tracking (as explained in Section~\ref{sec:consent_study}).
While the reported tracking domains can be used for tracking, they might also be used for other non-tracking purposes; however, it is the choice of the tracking company to designate domains for tracking.
We do not study the contents of network traffic because apps increasingly use certificate pinning (about $50\%$ of the studied apps used certificate pinning for some of their network communications).
As for our second study in Section~\ref{sec:developers}, we studied the online documentation of tracker libraries with great care, but did not always find all relevant information, particularly regarding the local storage of data on a user's device. Where this was the case, we disclosed this (e.g. see Table~\ref{tab:tracker_sdks}).

\section{Discussion and Future Work}
\label{sec:discussion}
Consent is an integral part of data protection and privacy legislation, both in the EU and the UK, and elsewhere.
This is all the more so in the context of third-party tracking, for which consent appears the only viable lawful ground
under the ePrivacy Directive and the GDPR, as analysed in Section~\ref{sec:consent_law}.
Not only has this has been emphasised by multiple DPAs, but is also acknowledged by tracking companies themselves in the documentation they make available to app developers.
Relying on the controller's legitimate interests---the only conceivable alternative to consent
under EU and UK data protection law---would likely fail short of passing the balancing test outlined in Article 6(1)f GDPR. 
This also follows from the requirement to obtain consent prior to storing or accessing information on a user's device,
under the ePrivacy Directive.

Against this backdrop, we analysed 1,297 mobile apps from Google Play in Section~\ref{sec:consent_study} and discovered a widespread lack of appropriate mechanisms to gather consent as required under the applicable regulatory framework.
We found that, while the guidelines of many commonly used tracker libraries require consent from EU and UK users, most apps on the Google Play Store that include third-party tracking features do not implement any type of consent mechanism.
The few apps that require data subjects to consent do so with regard to personalised advertising, but rarely for analytics---despite this being one of the most common tracking practices. 
Where an opt-out from personalised advertising was possible, the number of tracker domains contacted decreased only slightly after opting-out, hinting to continued data collection when serving contextual advertising.
These observations are at odds with the role of consent as the only viable option to justify the processing of personal data inherent to third-party tracking.

As detailed in Section~\ref{sec:consent_study}, the fact that only 9.9\% of the investigated apps request any form of consent already suggests widespread violations of current EU and UK privacy law.
This is even before considering the validity of the consent mechanisms put in place by that small fraction of apps.
As underlined in Section~\ref{sec:consent_law}, consent must be \enquote{freely given}, \enquote{informed}, \enquote{specific} and \enquote{unambiguous}.
The findings outlined in Section~\ref{sec:consent_study} suggest that most apps that do implement consent force users to grant consent, therefore ruling out its qualification as \enquote{freely given}.
The same goes for the 43.7\% of those apps that do not provide data subjects with the possibility to consent separately for each purpose, but instead rely on \emph{bulk} consent for a wide array of purposes.

When considering both the absence of any form of consent in more than 90\% of the investigated apps and the shortcomings inherent to the few consent mechanisms that are implemented by the remaining sample, we infer that the vast majority of mobile apps fail short of meeting the requirements stemming from EU and UK data protection law.
Our analysis does not even consider the fact that consent is only one of a variety of legal rules that third-party tracking needs to comply with.
Breaches of other legal principles---such as data minimisation, purpose and storage limitation, security and transparency---might be less visible than a lack of consent and harder to analyse, but no less consequential.

We further found that one of the reasons for the lack of consent implementation in apps might be inadequate support by tracker companies~\cite{balebako_privacy_2014,mhaidli_we_2019}.
Studying the online documentation of the 13 most commonly used tracker libraries in Section~\ref{sec:developers}, only 3 trackers implemented consent by default, and another 3 disclosed the need to implement consent as part of step-by-step implementation guides.
These step-by-step guides serve as a primary resource for app developers, and can give a false impression of completeness when in fact additional code needs to be added for many trackers to retrieve user consent.
This is true for at least 6 out 13 trackers, including Google Analytics and the Facebook App Events SDK, which likely need consent, but neither disclose this in their implementation guides nor implement such consent by default.
While most trackers provide some compliance guidance, we found that this can be difficult to find, hard to read, and poorly maintained.
Whatever the reasons for the lack of consent, the result is an absence of end-user controls for third-party tracking in practice.

Lastly, it is worth highlighting that Google, which is both the largest tracking company and the main developer of Android, faces conflicts of interest with respect to protecting user privacy in its Google Play ecosystem~\cite{greene_platform_2018,edelman2016android,cma_digital_markets2020}.
The company generates most of its revenue from personalised advertising, and relies on tracking individuals at scale.
Certain design choices by Google, including its ban of anti-tracking apps from the Play Store, its recent action against modified versions of Android, and the absence of user choice over AdID access for analytics on Android (as opposed to iOS), create friction for individuals who want to reduce data collection for tracking purposes, and lead to increased collection of personal data, some of which is unlawful as our legal analysis has shown.

\textbf{Future work.} An overarching question for future work is the extent of the legal obligations faced by the many actors involved in the third-party tracking ecosystem, ranging from app developers to providers of tracker libraries and mobile operating systems.
This is inextricably linked to their qualification as \enquote{controllers}, a legal notion the boundaries of which remain, despite recent jurisprudence~\cite{ecj_jw,ecj_waka,ecj_fashion} and detailed guidance~\cite{wp29_controllership,edpb_controllership}, still controversial.
Our analysis highlighted how simple changes in the software design can have significant effects for user privacy.

Moreover, while the US---unlike many developed countries---lack a federal privacy law, there exists a variety of specific privacy laws, such as COPPA to protect children and HIPAA to protect health data, as well as state-level privacy laws, including CCPA in California.
Some of these laws foresee consent requirements similar to EU and UK law.
We leave it to further work to assess how widely apps comply with the consent requirements of US privacy legislation.

\section{Conclusions}
\label{sec:conclusions}
Our work analyses the legal requirements for consent to tracking in apps, and finds an absence of such consent in practice based on an analysis of a representative sample of Google Play apps. This, in turn, suggests widespread violations of EU and UK privacy law. Simple changes by software intermediaries (such as Google and Facebook), including default consent implementations in tracker libraries, better legal guidance for app developers, and better privacy options for end-users, could improve the status quo around app privacy significantly.
However, there is doubt that these changes will happen without further intervention by independent parties---not only end-users, but also policymakers and regulators---due to inherent conflicts between user privacy and surveillance capitalism.

While the web has seen a proliferation of deceptive and arguably meaningless consent banners in recent years~\cite{matte_cookie_2019,nouwens_dark_2020}, we hope that mobile apps will not see a similar mass adoption. Rather, we aim to influence the current policy discourse around user choice over tracking and ultimately to make such choice more meaningful. As Apple has demonstrated with its recently introduced iOS 14.5, system-level user choices can
standardise the process of retrieving user consent and
make hidden data collection, such as tracking, more transparent to end-users.
We call on policy makers and data protection regulators to provide more stringent guidelines as to how consent to tracking should be implemented in program code, particularly by browser developers, device manufacturers, and platform gatekeepers in the absence of such existing requirements.

\bibliographystyle{plain}
\bibliography{bibliography}

\end{document}